\documentclass[conference]{IEEEtran}
\IEEEoverridecommandlockouts
\usepackage{cite}
\usepackage{amsmath,amssymb,amsfonts}
\usepackage{algorithmic}
\usepackage{graphicx}
\usepackage{textcomp}
\usepackage{xcolor}

\usepackage{multirow}
\usepackage{booktabs}

\usepackage{tikz}

\usepackage[colorlinks=true, urlcolor=blue, linkcolor=red]{hyperref}

\def\BibTeX{{\rm B\kern-.05em{\sc i\kern-.025em b}\kern-.08em
    T\kern-.1667em\lower.7ex\hbox{E}\kern-.125emX}}
\begin{document}

\title{Topo-VM-UNetV2: Encoding Topology into \\Vision Mamba UNet for Polyp Segmentation%\\
}

\author{
\IEEEauthorblockN{
   Diego Adame\IEEEauthorrefmark{1},
   Jose A. Nunez\IEEEauthorrefmark{1},
   Fabian Vazquez\IEEEauthorrefmark{1},
   Nayeli  Gurrola\IEEEauthorrefmark{1},
   Huimin Li\IEEEauthorrefmark{2},\\
   Haoteng Tang\IEEEauthorrefmark{1},
   Bin Fu\IEEEauthorrefmark{1},
   Pengfei Gu\IEEEauthorrefmark{1}
}
\IEEEauthorblockA{\IEEEauthorrefmark{1}Department of Computer Science, University of Texas Rio Grande Valley, Edinburg, TX 78539, USA}
\IEEEauthorblockA{\IEEEauthorrefmark{2}Department of
Mathematical Sciences, The University of Texas at Dallas, Richardson, TX
75080, USA}
}
% \author{
% \IEEEauthorblockN{
%    ******
% }
% \IEEEauthorblockA{******
% }
% }

\maketitle
\begin{abstract}
Convolutional neural network (CNN) and Transformer-based architectures are two dominant deep learning models for polyp segmentation. However, CNNs have limited capability for modeling long-range dependencies, while Transformers incur quadratic computational complexity. Recently, State Space Models such as Mamba have been recognized as a promising approach for polyp segmentation because they not only model long-range interactions effectively but also maintain linear computational complexity. However, Mamba-based architectures still struggle to capture topological features (e.g., connected components, loops, voids), leading to inaccurate boundary delineation and polyp segmentation. To address these limitations, we propose a new approach called Topo-VM-UNetV2, which encodes topological features into the Mamba-based state-of-the-art polyp segmentation model, VM-UNetV2. Our method consists of two stages: Stage 1: VM-UNetV2 is used to generate probability maps (PMs) for the training and test images, which are then used to compute topology attention maps. Specifically, we first compute persistence diagrams of the PMs, then we generate persistence score maps by assigning persistence values (i.e., the difference between death and birth times) of each topological feature to its birth location, finally we transform persistence scores into attention weights using the sigmoid function.
Stage 2: These topology attention maps are integrated into the semantics and detail infusion (SDI) module of VM-UNetV2 to form a topology-guided semantics and detail infusion (Topo-SDI) module for enhancing the segmentation results. 
Extensive experiments on five public polyp segmentation datasets demonstrate the effectiveness of our proposed method. The code will be made publicly available.
\end{abstract}

\begin{IEEEkeywords}
Topology, persistent homology, attention mechanism, deep learning, vision mamba, polyp segmentation
\end{IEEEkeywords}

\section{Introduction} \label{intro}
% 
%\vspace{-3mm}
Colorectal cancer (CRC) is one of the most common and deadly cancers worldwide, necessitating early intervention and accurate diagnosis for effective treatment and patient management~\cite{sung2021global,zitnik2024current,vazquez2025exploring,gu2024boosting}. Polyp segmentation is a critical task in CRC screening, aiming to precisely delineate the boundaries of polyps in endoscopic images to facilitate timely interventions by clinicians~\cite{fan2020pranet,zhang2024testfit}. Accurate polyp segmentation is central to reducing CRC-related mortality, as it enables colonoscopic polypectomy and lowers false-negative rates in screening tests~\cite{brandao2017towards,zhang2023swipe,zhang2023samdsk}.

%%%%%%%%%%%%%%%%%%%%%%%%%%%%%%%%%%%%%%
\begin{figure}[t]
\centering
\includegraphics[width=0.45\textwidth]{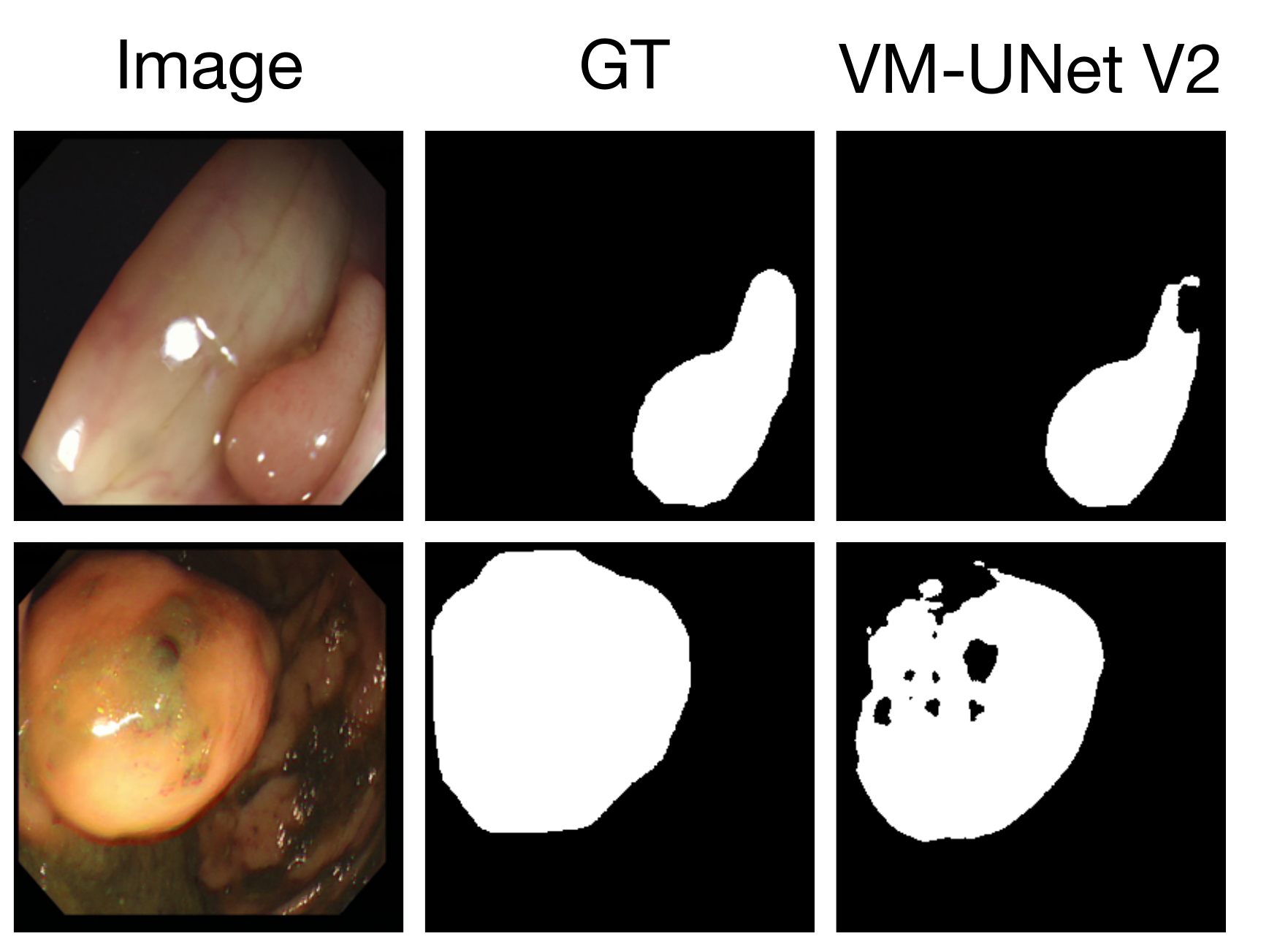}
\caption{Examples of polyp segmentations generated by VM-UNetV2~\cite{zhang2024vm}, showing that topological features (e.g., connected components) are not well captured.}
\label{fig:false_positives}
\end{figure}
%%%%%%%%%%%%%%%%%%%%%%%%%%%%%%%%%%%%%%

%%%%%%%%%%%%%%%%%%%%%%%%%%%%%%%%%%%%%%
\begin{figure*}[t]
\centering
\includegraphics[width=1.0\textwidth]{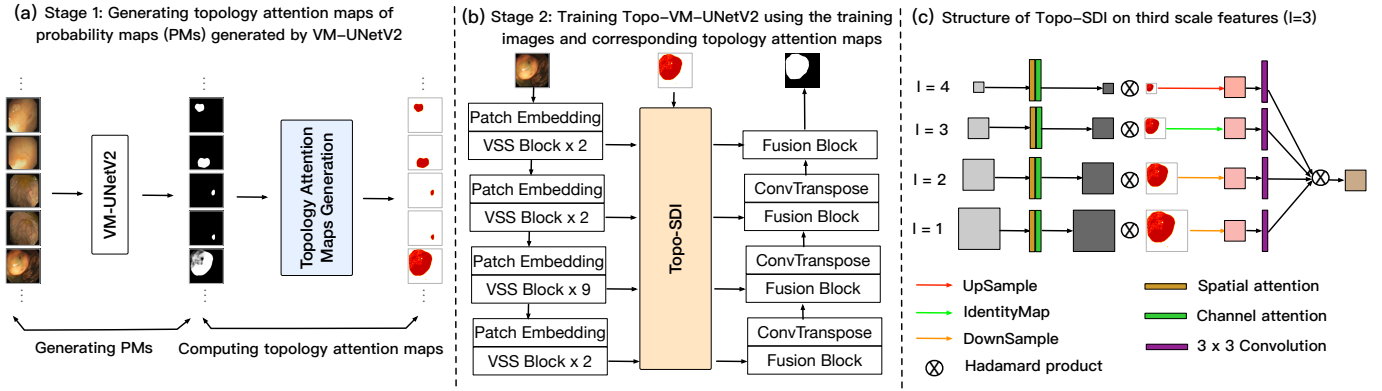}
\caption{The pipeline of our proposed framework. (a) Stage 1: VM-UNetV2~\cite{zhang2024vm} is used to generate probability maps for the training and test images, which are then used to produce corresponding topology attention maps. (b) Stage 2: The training images and their topology attention maps are used to train our Topo-VM-UNetV2 model, which consists of an Encoder, the Topo-SDI (topology-guided semantics and detail infusion) module, and a Decoder. (c) The structure of the Topo-SDI module. For simplicity, only the refinement of the third scale features ($l=3$) is shown.}
\label{fig:pipeline}
\end{figure*}
%%%%%%%%%%%%%%%%%%%%%%%%%%%%%%%%%%%%%%

Deep learning (DL) has significantly enhanced automated polyp segmentation performances. Convolutional neural network (CNN) models (e.g.,~\cite{gu2021k}), especially U-Net\cite{ronneberger2015u} variants (e.g., PraNet\cite{fan2020pranet} and SANet\cite{wei2021shallow}), have dominated the field, thanks to their strong feature-extraction capabilities and encoder-decoder design. However, CNN-based architectures often struggle to capture long-range dependencies due to limited receptive fields\cite{dosovitskiy2020image,gu2023convformer}.
To address this, Transformer-based architectures have been introduced to capture global contextual information, yielding substantial improvements in polyp segmentation accuracy (e.g., Polyp-PVT\cite{dong2021polyp} and U-Net v2\cite{peng2023u}). Nonetheless, these architectures are data-hungry and can incur quadratic computational complexity.
Recently, State Space Models (SSMs), particularly Structured SSMs (S4), have emerged as an effective solution, due to their proficiency in handling long sequences (e.g., Mamba\cite{gu2023mamba}). The addition of a cross-scan module in the Visual State Space Model (VMamba~\cite{liu2024vmamba}) has further strengthened Mamba’s suitability for polyp segmentation task, giving rise to models such as VM-UNet\cite{ruan2024vm} and VM-UNetV2\cite{zhang2024vm}.

Even with the enhanced performance of CNN, Transformer, and Mamba-based techniques, these models often remain unsatisfactory in real-world applications. One primary challenge is under- or over-segmentation: Normal tissues may be mistakenly recognized as polyps, or actual polyps may be missed, potentially leading to unnecessary surgeries or procedures. Moreover, these architectures often overlook essential topological aspects of polyps, such as boundary smoothness and global structure, leading to inaccurate boundary delineation and polyp segmentation. Fig.~\ref{fig:false_positives} shows polyp segmentations generated by a state-of-the-art (SOTA) model, VM-UNetV2~\cite{zhang2024vm}, emphasizing the need for precise boundary delineation and the capturing of relationships among regions and global structural patterns through well-captured topological features (e.g., connected components).

To address these limitations, we propose Topo-VM-UNetV2, a novel approach that incorporates topological features into the SOTA polyp segmentation model, VM-UNetV2. Our method consists of two stages: In Stage 1, we train VM-UNetV2 to generate probability maps (PMs) for both training and test images. These PMs are then used to compute topology attention maps. Specifically, we first compute persistence diagrams (PDs) from the PMs, then assign persistence values (the difference between death and birth times) of each topological feature to its birth location to obtain persistence score maps. Finally, we transform the persistence scores into attention weights via the sigmoid function, generating the topology attention maps. In Stage 2, these topology attention maps are incorporated into VM-UNetV2 to improve segmentation results. In particular, they are integrated into the semantics and detail infusion (SDI) module~\cite{peng2023u}, thus creating a topology-guided semantics and detail infusion (Topo-SDI) module that encodes influential topological information.

Our main contributions are three-fold: (1) We propose a new two-stage method to incorporate topological features into VM-UNetV2 by computing topology attention maps from probability maps.
(2) We develop a topology-guided semantics and detail infusion (Topo-SDI) module, which encodes the topological features to enhance polyp segmentation.
(3) Our method achieves SOTA performance on five public polyp segmentation datasets, surpassing existing SOTA methods.

%\vspace{-4mm}

%%%%%%%%%%%%%%%%%%%%%%%%%%%%%%%%%%%%%%
\begin{figure*}[t]
\centering
\includegraphics[width=1.0\textwidth]{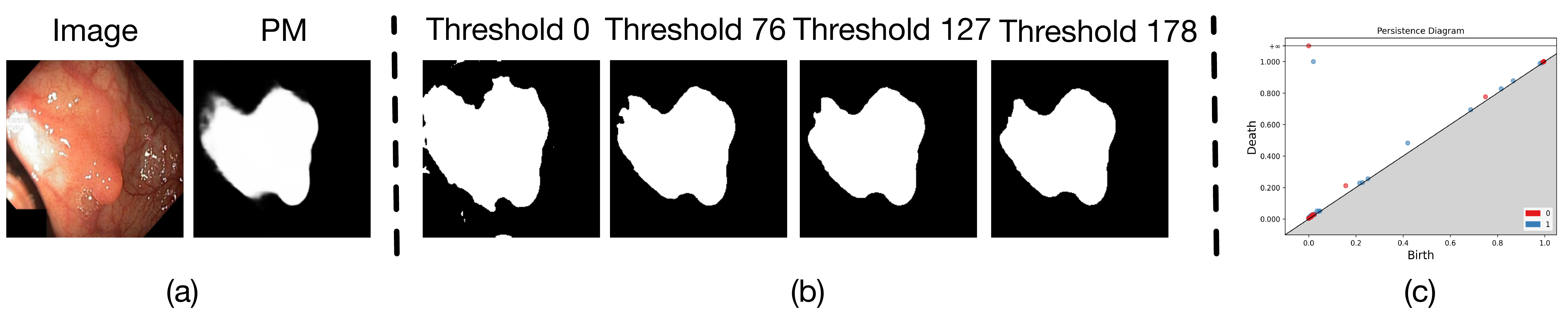}
\caption{Illustrating the sub-level filtration process. (a) An example image and its probability map generated by VM-UNetV2. 
(b) Four thresholded binary images of the probability map. As the threshold value increases or decreases, some connected components or loops are born or die. (c) The computed persistence diagram of the probability map, where red points denote 0-D persistent homology ($\text{PD}_0$) and blue points denote 1-D persistent homology ($\text{PD}_1$).}
\label{fig:sub-level}
\end{figure*}
%%%%%%%%%%%%%%%%%%%%%%%%%%%%%%%%%%%%%%

\section{Methods} 
Fig.~\ref{fig:pipeline} illustrates the pipeline of our Topo-VM-UNetV2, which comprises two main steps. First, we generate probability maps using VM-UNetV2 and compute the corresponding topology attention maps. Second, these topology attention maps are integrated into the SDI module of VM-UNetV2 to effectively capture topological features.

\subsection{Topology Attention Maps Generation}
\subsubsection{Review of VM-UNetV2} 
\textbf{Formulation of SSMs.}
SSMs, particularly Structured State Space Sequence models (S4) and Mamba~\cite{gu2023mamba}, have gained attention as efficient methods for modeling sequential information within DL architectures. These models interpret the input sequence as a continuous function \( x(t) \in \mathbb{R} \) that transforms through intermediate latent states \( h(t) \in \mathbb{R}^{N} \), resulting in an output \( y(t) \in \mathbb{R} \). This process can be represented mathematically by a linear ordinary differential equation:

\begin{equation}
\begin{aligned}
\frac{dh(t)}{dt} &= A h(t) + B x(t), \\
y(t) &= C h(t),
\end{aligned}
\end{equation}

where \( A \in \mathbb{R}^{N \times N} \) is a state transition matrix, and \( B, C \in \mathbb{R}^{N \times 1} \) represent the input and output projection parameters, respectively.

To adapt this continuous-time model for use in discrete DL frameworks, a discretization step is applied by introducing a timestep parameter, \( \Delta \). This step converts continuous parameters \( A \) and \( B \) into discrete counterparts \( \overline{A} \) and \( \overline{B} \). The widely used discretization approach, known as zero-order hold, is defined as follows:

\begin{equation}
\begin{aligned}
\overline{A} &= \exp(\Delta A), \\
\overline{B} &= (\Delta A)^{-1}\left(\exp(\Delta A) - I\right) \Delta B.
\end{aligned}
\end{equation}

After discretization, the resulting SSMs can be computed through two equivalent forms: (1) linear recurrence, where states are updated sequentially, or (2) global convolution, which applies a convolutional kernel simultaneously across the input sequence. Formally, these computations are represented by:

Linear recurrence:
\begin{equation}
\begin{aligned}
h'(t) &= \overline{A} h(t) + \overline{B} x(t), \\
y(t) &= C h(t),
\end{aligned}
\end{equation}

global convolution:
\begin{equation}
\begin{aligned}
K &= [C\overline{B}, C\overline{A}\,\overline{B}, \dots, C\overline{A}^{L-1}\overline{B}], \\
y &= x * K,
\end{aligned}
\end{equation}

where \( K \in \mathbb{R}^{L} \) denotes the structured convolution kernel derived from the discretized state-space parameters, \( L \) is the length of the input sequence, and \( * \) indicates the convolution operation.

\textbf{2D Selective-Scan (SS2D).}
The data forwarding process in SS2D consists of three main steps: \textit{cross-scan}, selective scanning with S6 blocks (SSM + Selection), and \textit{cross-merge}. Specifically, SS2D initially unfolds the input image patches into sequences along four distinct traversal paths (\textit{cross-scan}). Each resulting patch sequence is then independently processed by a dedicated S6 block. Subsequently, these processed sequences are reshaped and combined (\textit{cross-merge}) to generate the final output feature map. By leveraging complementary one-dimensional traversal paths, SS2D enables each pixel in the image to aggregate contextual information from multiple directions, thereby effectively constructing global receptive fields within the 2D spatial domain.

\textbf{Visual State Space (VSS) Block.}
Initially, the input passes through a linear embedding layer, after which it splits into two parallel processing streams. The first stream involves a $3\times 3$ depth-wise convolution followed by a SiLU activation function, feeding subsequently into the main SS2D module. The output from SS2D is then normalized via layer normalization before being fused with the second stream, which independently undergoes a SiLU activation. The resulting combined output represents the final feature map generated by the VSS block.

\textbf{Structure of VM-UNetV2.}
The architecture of VM-UNetV2 comprises three main components: An encoder, a SDI module, and a decoder. In the encoder, VSS blocks are organized into four hierarchical stages, with each stage containing [\(2, 2, 9, 2\)] blocks and corresponding feature dimensions of [\(96, 2*96, 4*96, 8*96\)]. The decoder consists of ConvTranspose layers and addition fusion blocks.

VM-UNetV2 is a SOTA polyp segmentation architecture (see Table~\ref{tab:polyp}), yet it does not effectively capture topological features such as connected components and holes. To address this limitation, we propose integrating topology attention maps, derived from probability maps, into the model. Here, we use VM-UNetV2 to generate the probability maps for both training and test images.

\subsubsection{Topology Attention Maps Generation}
Persistent Homology (PH) is a foundational mathematical tool in topological data analysis for analyzing topological properties of diverse data types, including point clouds, networks, and images~\cite{edelsbrunner2010computational}. It is capable of identifying and quantifying important features of data that persist over a range of different spatial scales, thus
providing insight into the underlying structures of the data. 

Here, we outline how PH is used to compute topology attention maps from probability maps through three main steps: (1) Constructing filtrations, (2) computing persistence diagrams, and (3) generating topology attention maps. For in-depth theoretical details, please refer to~\cite{edelsbrunner2010computational,dey2022computational}.

\textbf{Step 1: Constructing Filtrations.}
Filtration is the process of building a sequence of nested topological spaces that 
represent the progressive inclusion (or exclusion) of features in a dataset. In image-based  applications, these topological spaces are often built as \emph{cubical complexes}, topological structures derived from binary images.

Given a probability map $P$ (with dimension $256 \times 256$) where each pixel 
$\Delta_{ij} \subset P$ has an associated probability value $p_{ij}$. Here, we construct filtration in sub-level filtration process: Choose an increasing set of thresholds
  \[
    0 \;\le\; t_{1} \;<\; t_{2} \;<\; \dots \;<\; t_{N} \;\le\; 255.
  \]
  At each threshold $t_{n}$, we ``activate'' (color black) all pixels whose probability 
  is \emph{less than or equal to} $t_{n}$. Formally,
  \[
    X_{n} \;=\; \{\,\Delta_{ij}\subset P \;\mid\; p_{ij} \;\le\; t_{n}\}.
  \]
  As $t_{n}$ grows, more pixels become activated, yielding a nested sequence of binary images:
  \[
    X_{1} \;\subset\; X_{2} \;\subset\; \cdots \;\subset\; X_{N}.
\]
By converting each thresholded image into a binary mask and then 
forming its associated cubical complex, one can examine how topological features appear and vanish as thresholds change. Fig.~\ref{fig:sub-level} (b) illustrates an example of the sub-level filtration process.

\textbf{Step 2: Computing Persistence Diagrams (PDs).}
PH captures the evolution of topological features 
(e.g., connected components, loops) across a filtration of the 
\emph{probability map}. This information is recorded in a \emph{PD}, 
which maps each feature~$\sigma$ to its birth and death thresholds $(b_{\sigma}, d_{\sigma})$. 
For a sublevel filtration $\{X_n\}$ of a probability map, the \emph{birth time} $b_{\sigma}$ 
is the threshold $t_m$ at which the feature $\sigma$ first appears, and the \emph{death time} 
$d_{\sigma}$ is the threshold $t_n$ at which $\sigma$ disappears. 
The PD $\text{PD}_k(X)$ for dimension~$k$ collects all such pairs 
for $k$-dimensional features (e.g., $\text{PD}_0$ for connected components, $\text{PD}_1$ 
for loops). Features with long lifespans $(d_{\sigma} - b_{\sigma})$ are typically considered 
more significant, as they represent prominent topological structures. Fig.~\ref{fig:sub-level} (c) shows the computed PD of the probability map, where red points represent connected components and blue points indicate holes/loops.

\textbf{Step 3: Generating Topology Attention Maps.}
Although PDs are informative, they are composed of 2D points that are challenging to use directly in DL models. To make PDs more compatible with DL architectures, we convert them into topology attention maps through the following steps: (1) Significance Filtering: We calculate persistence values as the difference between death and birth times, then apply a threshold filter (at the 50th percentile) to retain only topologically significant features. (2) Birth Location Mapping: For each significant feature in the PDs, we locate cells (pixels) whose filtration values closely match the feature’s birth value. We assume that cells with values very close to the birth value correspond to the birth locations.
(3) Persistence Score Assignment: We create a persistence score map by assigning each birth location a persistence value. (4) Attention Weight Generation: We transform these persistence scores into attention weights by applying the sigmoid function, resulting in values within [0,1] that highlight topologically important regions.

Our topology attention maps maintain the same spatial dimensions (height and width) as the probability maps and original images, facilitating direct integration with existing DL models. But the question is how to incorporate these maps in a way that ensures the model focuses on structurally significant regions based on their topological persistence.

\subsection{Topology-guided Semantics and Detail Infusion (Topo-SDI) Module}
To effectively encode topological features into VM-UNetV2, we integrate the topology attention maps into the SDI module. As illustrated in Fig.~\ref{fig:pipeline} (c), given the multi-scale feature maps $f^0_i$ ($i = 1, 2, 3, 4$) generated by the encoder, we first feed them into the CBAM attention mechanism~\cite{woo2018cbam} to learn both local spatial information and global channel information, yielding: $f^1_i = \phi^{att}_i(f^0_i)$, where $\phi^{att}_i$ denotes the CBAM parameters at the $i$-th scale, following~\cite{peng2023u,zhang2024vm}. Next, we apply a $1\times 1$ convolution to reduce the number of channels in $f^1_i$ to $c$. The resulting feature map is denoted as $f^2_i\in R^{H_i\times W_i\times c}$, where $H_i$, $W_i$, and $c$ denote the height, width, and channel dimensions of $f^2_i$, respectively.

To enable the SDI module to capture topological features, we first resize the topology attention map to produce multi-scale topology attention maps $T_i$ ($i = 1, 2, 3, 4$), ensuring $T_i$ matches the dimensions of the feature map $f^2_i$. We then apply the element-wise Hadamard product ($H(.,.)$) between $f^2_i$ and $T_i$, enriching the $i$-th scale features with topological information:
\begin{eqnarray}
    f^3_i = H(f^2_{i}, T_i).
\end{eqnarray}

Next, we fuse the refined multi-scale feature maps $f^3_i$ ($i = 1, 2, 3, 4$) at the $i$-th scale via an element-wise Hadamard product before passing it to the $i$-th decoder. Specifically, using $f^3_i$ as the target reference, we first adjust the resolutions of the feature maps at each $j$-th level to match $f^3_i$, as follows:
\begin{equation}
f^4_{ij} =
\begin{cases}
\text{Conv}(\text{DownSample}(f^3_j, (H_i, W_i)))& \text{if $j < i$ },\\[5pt]
\text{Conv}(\text{IdentityMap}(f^3_j)) & \text{if $j=i$},\\[5pt]
\text{Conv}(\text{UpSample}(f^3_j, (H_i, W_i))) & \text{if $j>i$},
\end{cases}       
\end{equation}
where $\text{DownSample}$, $\text{IdentityMap}$, and $\text{UpSample}$ represent adaptive average pooling, identity mapping, and bilinear interpolation of $f^2_j$ to the resolution $H_i\times W_i$, respectively (with $1\leq i, j  \leq M$). $\text{Conv}$ denotes a $3\times 3$ convolution. Subsequently, $f_{i}^{5}=H\left (  f_{i1}^{4}, f_{i2}^{4},f_{i3}^{4}, f_{i4}^{4}   \right )$ is sent to the decoder at the $i$-th level for further resolution reconstruction and segmentation.

We refer to this new topology-guided semantics and detail infusion module as Topo-SDI. By replacing the original SDI module with Topo-SDI in VM-UNetV2, we obtain the updated architecture for polyp segmentation, as shown in Fig.~\ref{fig:pipeline} (b).

\begin{table}[t]
\centering
\caption{Experimental comparison of different methods on the five polyp segmentation datasets. The best results are highlighted in bold.}
% \small
\begin{tabular}{l|lllll}
Datasets & Methods & DSC (\%) & IoU (\%) & MAE  \\
\hline 
\multirow{6}{*}{Kvasir-SEG} & 
U-Net~\cite{ronneberger2015u}  &85.41	 &78.22	 &0.044  \\
& U-Net v2~\cite{peng2023u}  &89.93	&84.05	&0.029 \\
& VM-UNet~\cite{ruan2024vm}  &86.21	&79.13	&0.041 \\
& VM-UNetV2~\cite{zhang2024vm} & 90.75	&84.86	&0.027  \\\cline{2-5}
& Topo-VM-UNetV2 (ours)&\textbf{91.95} & \textbf{86.54} & \textbf{0.021} \\
\hline 

\multirow{6}{*}{ClinicDB} & 
U-Net~\cite{ronneberger2015u} &86.77	&81.11	&0.019 \\
& U-Net v2~\cite{peng2023u}&89.33	&84.25	&0.014 \\
& VM-UNet~\cite{ruan2024vm}  &87.11	&82.37	&0.019 \\
& VM-UNetV2~\cite{zhang2024vm} &90.07	&84.68	&0.016 \\\cline{2-5}
& Topo-VM-UNetV2 (ours)&\textbf{91.83} & \textbf{86.80} & \textbf{0.008} \\
\hline 

\multirow{6}{*}{ColonDB} & 
U-Net~\cite{ronneberger2015u} &65.41	&57.62	&0.045 \\
& U-Net v2~\cite{peng2023u} &77.37	&69.49	&0.039 \\
& VM-UNet~\cite{ruan2024vm}  &71.72	&62.16	&0.042 \\
& VM-UNetV2~\cite{zhang2024vm} &77.29	&69.19	&0.031  \\\cline{2-5}
& Topo-VM-UNetV2 (ours)&\textbf{79.00} & \textbf{70.55} & \textbf{0.028} \\
\hline 

\multirow{6}{*}{ETIS} & 
U-Net~\cite{ronneberger2015u} &54.85	&49.02	&0.022 \\
& U-Net v2~\cite{peng2023u} &68.98	&60.41	&0.029 \\
& VM-UNet~\cite{ruan2024vm}  &57.01	&48.21	&0.046 \\
& VM-UNetV2~\cite{zhang2024vm} &72.39	&63.84	&0.017  \\\cline{2-5}
& Topo-VM-UNetV2 (ours)&\textbf{75.68} & \textbf{66.72} & \textbf{0.014} \\
\hline 

\multirow{6}{*}{CVC-300} & 
U-Net~\cite{ronneberger2015u}&82.38	&74.69	&0.010 \\
& U-Net v2~\cite{peng2023u} &86.70	&79.80	&0.010 \\
& VM-UNet~\cite{ruan2024vm}  &83.60	&74.43	&0.012\\
& VM-UNetV2~\cite{zhang2024vm} &87.14	&79.63	&0.008  \\\cline{2-5}
& Topo-VM-UNetV2 (ours)&\textbf{89.39} & \textbf{81.98} & \textbf{0.006} \\
\end{tabular}
\label{tab:polyp}
\end{table}
%%%%%%%%%%%%%%%%%%%%%%%%%%%%%%%%%%%%%%

\section{Experiments and Results} \label{exp}
\label{sec:exp}

\subsection{Datasets}
Five polyp segmentation datasets are used in the experiments: Kvasir-SEG~\cite{jha2020kvasir}, ClinicDB~\cite{bernal2015wm}, ColonDB~\cite{tajbakhsh2015automated}, ETIS~\cite{silva2014toward}, and CVC-300~\cite{vazquez2017benchmark}. For fair comparison, we followed the train/test split strategy in~\cite{peng2023u,zhang2024vm}. Specifically, 900 images from Kvasir-SEG and 550 images from ClinicDB are used as the training set. The test set includes all images from CVC-300 (60 images), ColonDB (380 images), and ETIS (196 images), as well as the remaining 100 images from Kvasir-SEG and the remaining 62 images from ClinicDB.

\subsection{Experimental Setup}
Our experiments are conducted using the PyTorch and MONAI. The model is trained on an NVIDIA Tesla V100 Graphics Card (32GB GPU memory) using the AdamW optimizer with a weight decay = 0.005. Following~\cite{zhang2024vm}, we resize all the images of each dataset to $256 \times 256$, set the learning rate to 0.001, batch size to 80, and the number of training epochs to 300 for the experiments. We make use of CosineAnnealingLR as the scheduler, with its operation spanning a maximum of 50 iterations and the learning rate going as low as 1e-5. Standard data augmentation (e.g., random flipping and random rotation) is applied to avoid overfitting. The encoder units’ weights are initially set to align with those of VMamba-S~\cite{liu2024vmamba}. We report DSC (Dice Similarity Coefficient) and IoU (Intersection over Union) scores, and MAE (Mean Absolute Error) scores. Each experiment is run 5 times using different seeds, and the averaged results are reported.

\subsection{Experimental Results}
Table~\ref{tab:polyp} presents a quantitative comparison of various methods on five polyp segmentation datasets. From these results, we observe the following: (1) Our approach achieves SOTA results on all five datasets. Note that we do not compare against PraNet~\cite{fan2020pranet}, SANet~\cite{wei2021shallow}, and Polyp-PVT~\cite{dong2021polyp}, as U-Net v2~\cite{peng2023u} has already demonstrated superior performance to these models on these five datasets. (2) Our method outperforms the previous SOTA model VM-UNetV2 on all metrics, with an improvement of up to 3.29\% in the DSC score on the ETIS dataset. These results verify the effectiveness of our method.

\begin{figure}[h!]
\centering
\includegraphics[width=0.5\textwidth]{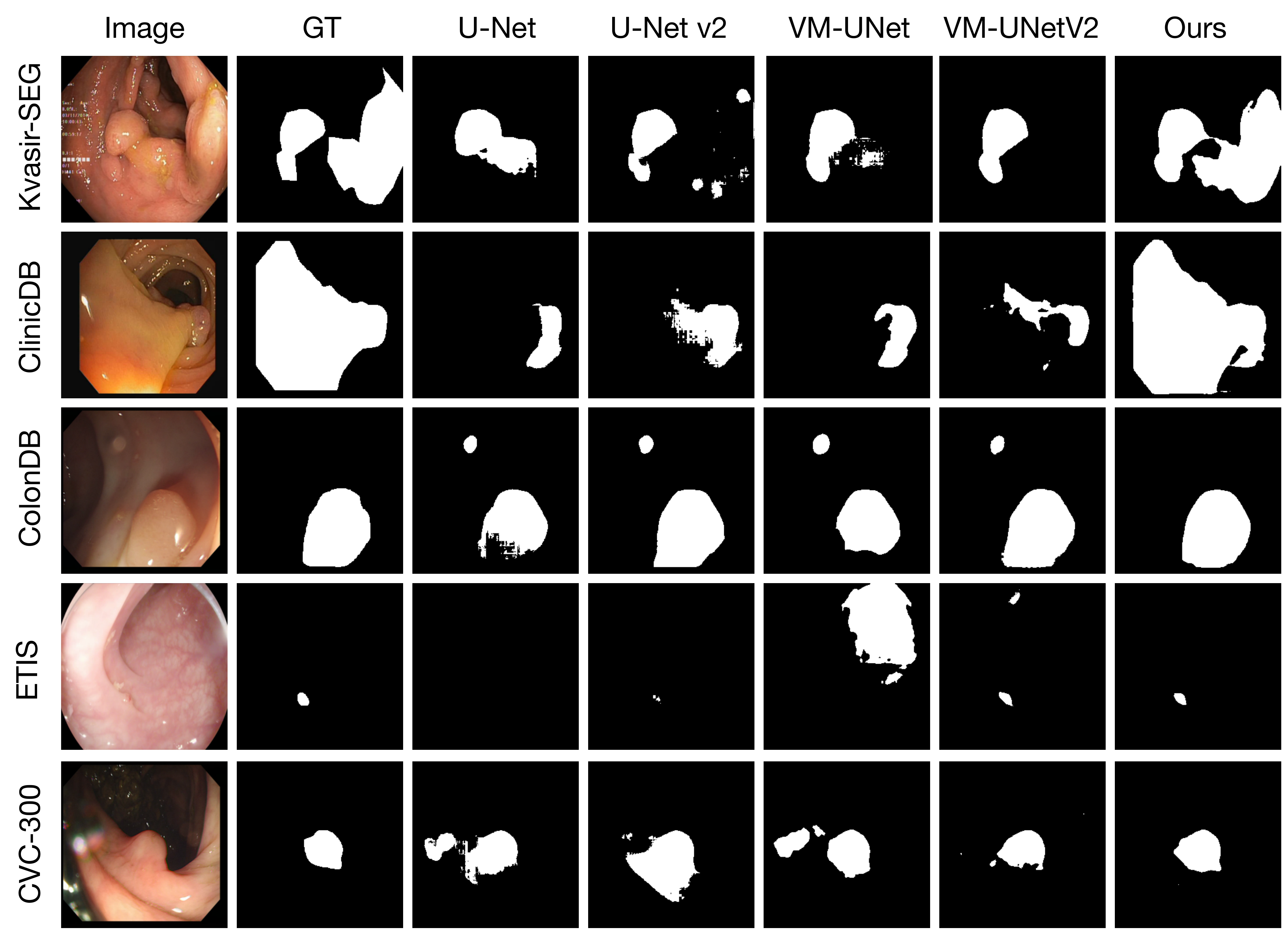}
\caption{Examples of segmentations from five polyp segmentation datasets.}
\label{fig:visual-results}
\end{figure}

Fig.~\ref{fig:visual-results} illustrates several segmentation examples from the five datasets. From these visual results, we observe the following: (1) The segmentation results produced by our method are significantly better than those of the other methods (U-Net, U-Net v2, VM-UNet, and VM-UNetV2). (2) Compared with VM-UNetV2, our results are closer to the ground truth (GT) across all five datasets. In particular, our method effectively corrects false positive errors (as shown in ColonDB and ETIS) and refines boundaries (as shown in CVC-300). These results confirm the effectiveness of our approach for encoding topological features.

\subsection{Ablation Study}
To evaluate the impact of the computed topology attention maps in the Topo-SDI module, we perform an ablation study on all five datasets. We consider two attention mechanisms within the SDI module: (1) PM-SDI: Probability maps are directly used to compute attention weights via the sigmoid function, and these weights serve as the topology attention maps in the Topo-SDI. (2) Topo-SDI: Topology attention maps are integrated into the SDI module. As shown in Table~\ref{tab:ablation}, PM-SDI yields slight improvements over VM-UNetV2 on Kvasir-SEG, ClinicDB, ColonDB, and CVC-300 datasets, while Topo-SDI achieves significantly greater gains across all five datasets.

\begin{figure}[h!]
\centering
\includegraphics[width=0.5\textwidth]{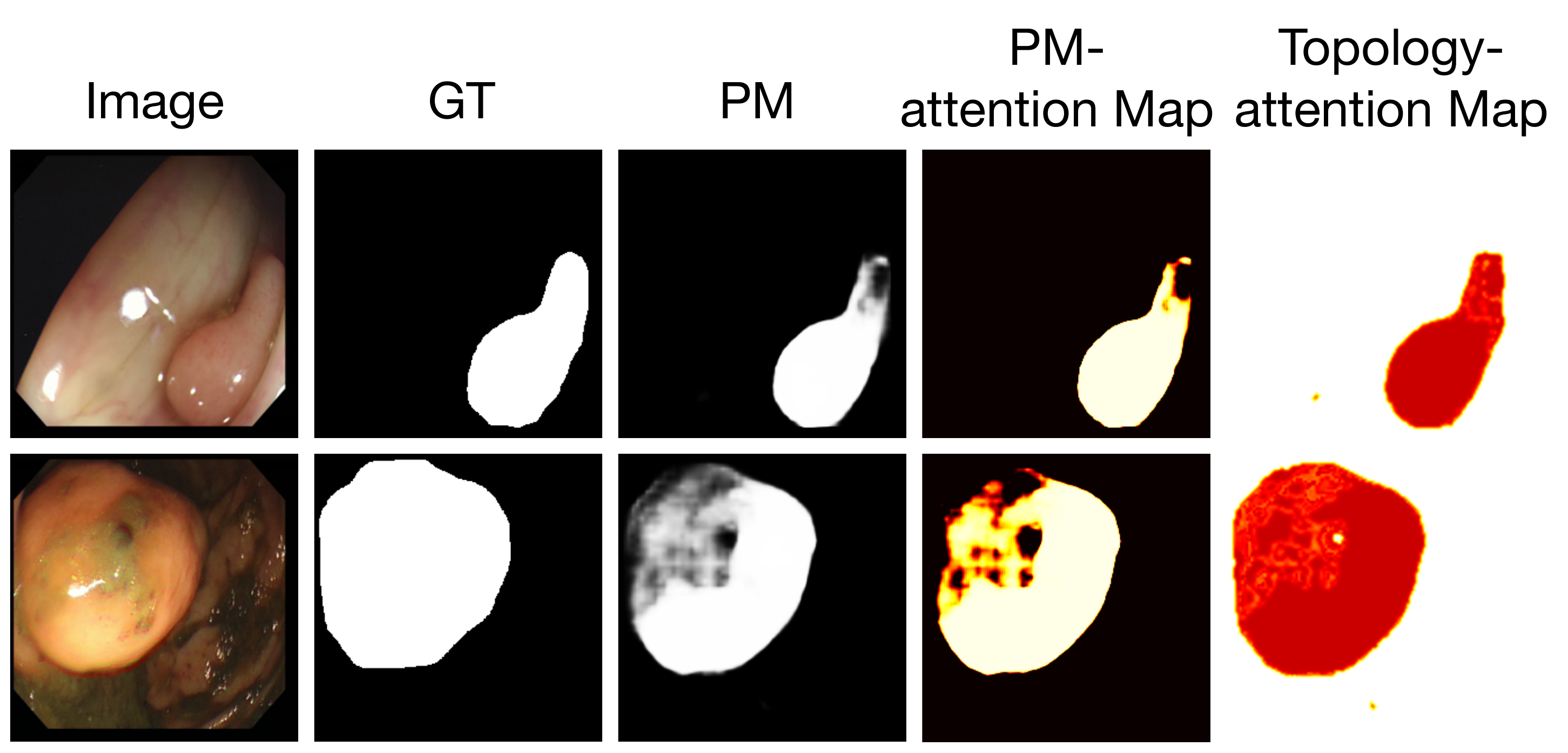}
\caption{Examples of different attention maps computed from probability maps.}
\label{fig:visual-attentions}
\end{figure}

Fig.~\ref{fig:visual-attentions} illustrates several examples of attention maps computed from probability maps. We observe the following: (1) Compared with the attention maps computed directly from probability maps, topology attention maps effectively emphasize the boundaries between different tissue types, resulting in more accurate boundary delineation. (2) Topology attention maps capture relationships between regions and global structural patterns, highlighting areas that preserve the overall topology. These observations suggest that probability attention maps primarily emphasize the centers of high-probability regions, whereas topology attention maps dedicate more focus to boundaries and transitions between regions, key to maintaining topological structure. Consequently, topology attention maps encode richer topological information than raw probability values alone.

%%%%%%%%%%%%%%%%%%%%%%%%%%%%%%%%%%%%%%
%\setlength{\tabcolsep}{3pt}
\begin{table}[t]
\centering
% \small
\caption{Ablation study on the five polyp datasets.}
\begin{tabular}{l|llll}
Dataset & Method  & DSC (\%)& IoU (\%) \\
% \cline{2-5}
\hline 
& VM-UNetV2~\cite{zhang2024vm} & 90.75	&84.86  \\
\multirow{1}{*}{Kvasir-SEG} 
& VM-UNetV2 + PM-SDI &90.77	&85.12 \\
%& VM-UNetV2 + Topo-SDI &  & \\
\cline{2-5}
& VM-UNetV2 + Topo-SDI  (ours) &\textbf{91.95} & \textbf{86.54} \\
\hline 
& VM-UNetV2~\cite{zhang2024vm} &90.07	&84.68 \\
\multirow{1}{*}{ClinicDB} 
& VM-UNetV2 + PM-SDI &91.04	&85.62 \\
%& VM-UNetV2 + Topo-SDI &  & \\
\cline{2-5}
& VM-UNetV2 + Topo-SDI  (ours) &\textbf{91.83} & \textbf{86.80} \\\hline
& VM-UNetV2~\cite{zhang2024vm} &77.29	&69.19  \\
\multirow{1}{*}{ColonDB} 
& VM-UNetV2 + PM-SDI &78.09	&69.58 \\
%& VM-UNetV2 + Topo-SDI &  & \\
\cline{2-5}
& VM-UNetV2 + Topo-SDI  (ours) &\textbf{79.00} & \textbf{70.55} \\
\hline 
& VM-UNetV2~\cite{zhang2024vm}&72.39	&63.84  \\
\multirow{1}{*}{ETIS} 
& VM-UNetV2 + PM-SDI &72.31	&62.91  & \\
%& VM-UNetV2 + Topo-SDI &  & \\
\cline{2-5}
& VM-UNetV2 + Topo-SDI  (ours) &\textbf{75.68} & \textbf{66.72} \\\hline 
& VM-UNetV2~\cite{zhang2024vm} &87.14	&79.63  \\
\multirow{1}{*}{CVC-300} 
& VM-UNetV2 + PM-SDI &87.21	&79.62   \\
%& VM-UNetV2 + Topo-SDI &  & \\
\cline{2-5}
& VM-UNetV2 + Topo-SDI  (ours) &\textbf{89.39} & \textbf{81.98} \\
\end{tabular}
\label{tab:ablation}
\end{table}
%%%%%%%%%%%%%%%%%%%%%%%%%%%%%%%%%%%%%%

%%%%%%%%%%%%%%%%%%%%%%%%%%%%%%%%%%%%%%
%%%%%%%%%%%%%%%%%%%%%%%%%%%%%%%%%%%%%%

%%%%%%%%%%%%%%%%%%%%%%%%%%%%%%%%%%%%%%
%%%%%%%%%%%%%%%%%%%%%%%%%%%%%%%%%%%%%%

\subsection{Computational Complexity of Topology Attention Maps}
Our topology attention maps computation is performed on an NVIDIA Tesla V100 GPU with 32GB of memory, using the Gudhi package. We report the time required to process the training set: Computing the topology attention maps for 1450 images (each sized $256 \times 256$ pixels) takes 186.99 seconds in total.

%\vspace{-4mm}

\section{Conclusions} \label{concl}
In this paper, we proposed Topo-VM-UNetV2, a new approach that encodes topological features into the Mamba-based SOTA polyp segmentation model, VM-UNetV2. Our method first computes topology attention maps from probability maps, which are then integrated into the SDI module of VM-UNetV2, resulting in a topology-guided semantics and detail infusion (Topo-SDI) module. Experiments on five public datasets demonstrated the effectiveness of our proposed method for polyp segmentation.

\section{Acknowledgement}
\label{sec:acknowledgements}
This research was supported in part by the Graduate Assistance in Areas of National Need (GAANN) grant.

\bibliographystyle{IEEEtran} 

\bibliography{ref}
\end{document}